\documentclass[runningheads]{llncs}

\usepackage{url}
\usepackage{float}
\usepackage{cite}
\usepackage{amsmath,amssymb,amsfonts}
\usepackage{algorithmic}
\usepackage{graphicx}
\usepackage{subcaption}
\usepackage{textcomp}
\usepackage{xcolor}

\usepackage{hyperref}
\hypersetup{breaklinks}

\usepackage{tikz}
\usepackage{bbold}
\usepackage[shortlabels]{enumitem}
\usepackage[normalem]{ulem}

\begin{document}

    \newboolean{showcomments}
    \setboolean{showcomments}{true}
    \ifthenelse{\boolean{showcomments}}
    {\newcommand{\mynote}[3] {
        \fbox{\bfseries\sffamily\scriptsize#1}
        {\small$\blacktriangleright$\textcolor{#3}{\textbf{#2}}}$\blacktriangleleft$
    }}

    \newcommand{\Anto}[1]{\mynote{Anto}{#1}{magenta}}
    \newcommand{\Tiphaine}[1]{\mynote{Tiphaine}{#1}{blue}}
    \newcommand{\Sara}[1]{\mynote{Sara}{#1}{red}}

    \title{A Comparative Gas Cost Analysis of Proxy and Diamond Patterns in EVM Blockchains for Trusted Smart Contract Engineering}
    \titlerunning{Gas Cost Analysis of Proxy and Diamond Patterns}

    \author{Anto Benedetti\inst{1,2}\orcidID{0009-0009-5526-1287}\and\\
    Tiphaine Henry\inst{1}\orcidID{0000-0002-7981-8934}
    \and\\
    Sara Tucci-Piergiovanni\inst{1}\orcidID{0000-0001-9738-9021}
    }
    \authorrunning{A. Benedetti, T.Henry, and S.Tucci-Piergiovanni}

    \institute{Université Paris-Saclay, CEA, List, F-91120, Palaiseau, France\\
    \and École Supérieure de Génie Informatique, Paris, France}

    \maketitle

    \begin{abstract}
        Blockchain applications are witnessing rapid evolution, necessitating the integration of upgradeable smart contracts.
        Software patterns have been proposed to summarize upgradeable smart contract best practices.
        However, research is missing on the comparison of these upgradeable smart contract patterns, especially regarding gas costs related to deployment and execution.
        This study aims to provide an in-depth analysis of gas costs associated with two prevalent upgradeable smart contract patterns: the proxy and diamond patterns.
        The proxy pattern utilizes a proxy pointing to a logic contract, while the diamond pattern enables a proxy to point to multiple logic contracts.
        A comparative analysis of gas costs for both patterns is conducted and compared to a traditional non-upgradeable smart contract.
        From this analysis, a theoretical contribution is derived in the form of two consolidated blockchain patterns and a corresponding decision model.

    \end{abstract}

\section{Introduction}\label{sec:introduction}

Smart contracts are pivotal for orchestrating digital transactions in a reliable and secure manner in blockchain platforms~\cite{kannengiesser2021challenges}.
Smart contracts are particularly important on Ethereum and similar blockchain platforms, where they are used in diverse areas including digital finance or industrial traceability.

As blockchain applications evolve, the need for smart contracts to be not only secure but also upgradeable becomes apparent~\cite{chen2020maintaining}.
In this context, a set of upgradeable smart contract patterns, including the proxy and diamond patterns, have surfaced to answer the lack of classical smart contracts' adaptability whose logic cannot be changed once deployed~\cite{meisami2023comprehensive,qasse2023smart}.
The proxy pattern uses a proxy contract to delegate calls to an implementation contract, providing a flexible and upgradeable solution.
The diamond pattern, introduced in EIP-2535, addresses concerns like contract size limitations and facilitates enhanced maintainability and versioning through multiple implementation contracts.

Research has focused on identifying upgradeable smart contracts pattern families, pointing towards proxy and diamond-based strategies ~\cite{marchesi2020design, ebrahimilarge, meisami2023comprehensive, qasse2023smart}.
However, the papers do not provide an in-depth analysis of the functional and non-functional properties of these patterns, nor gas costs behaviors.
Hence, a research gap is identified regarding a thorough study of the proxy and diamond patterns, especially in terms of a gas costs analysis.

To address this issue, this paper aims to answer the following research questions: \textit{(RQ1) How do the classic, proxy, and diamond patterns differ in terms of gas consumption, scalability, and ease of use?} And \textit{(RQ2) What implications do these differences have for the development of blockchain applications, considering the traditional classic pattern as a baseline?}

In this paper, we contribute to the literature through a unified approach to compare gas costs in upgradeable smart contracts.
We leverage this methodology to provide a comparative gas cost analysis of the proxy and diamond patterns in EVM blockchains, compared against a monolithic non-upgradeable smart contract, which is used as a baseline.
Based on these results, we derive a theoretical contribution in the form of two smart contract patterns adhering to the Alexandrian form format following the standard proposed by Christopher Alexander~\cite{alexander1977pattern}.
These patterns include the results of our comparative analysis
and contribute to the broader understanding of upgradeable smart contract patterns and their use in blockchain applications.
Based on these patterns, a decision model for using these patterns is proposed, emphasizing functional and non-functional properties for each design decision.

The remainder of this paper is structured as follows.
Section~\ref{sec:background} introduces key concepts related to smart contracts and blockchain patterns, and section~\ref{sec:related-work} presents studies already made on linked concepts.
Section~\ref{sec:methodology} presents the method used to carry a comparative analysis on proxy versus diamond gas costs.
Section~\ref{sec:evaluation} presents the results of the tests made on the different patterns.
Section~\ref{sec:discussion} leverages these findings to propose two consolidated proxy and diamond patterns as well as a decision model for using these patterns.
Section~\ref{sec:conclusion} finally concludes the paper with a summary and a discussion of the results and some considerations for future work.

\section{Background}\label{sec:background}

\subsection{Smart Contracts}\label{subsec:smart-contracts}

A smart contract is a program hosted on a blockchain network~\cite{szabo1997formalizing}.
When a smart contract executes, its updated state (or storage) is registered into a transaction.
Then, that transaction is stored in the blockchain ledger making it immutable and tamper-proof~\cite{ayub2023storage}.
More precisely, the final validated results are stored in a Merkle Patricia trie whose nodes correspond to an account or a smart contract.
A smart contract comprises both variables and functions.
Smart contract functions can execute arbitrary code, access the state of the variables and optionally update them.
The default function's mutability gives the right to read and modify state variables.
However, function mutability can be constrained to add more security when it comes to accessing these variables.
On the one hand, a \verb|pure| function cannot read nor modify the state of the contract; it is only used to compute a value, often using the parameters passed to it. On the other hand, a \verb|view| function can read the state of the contract, but cannot modify it. Smart contracts are immutable, meaning that once deployed, they cannot be modified~\cite{wood2014ethereum,buterin2017ethereum}.
This is a security feature, as it prevents malicious actors from modifying the code, so that the contract can be trustable and unbreakable.
However, this also means that if a bug is found in the code, it cannot be fixed, and the only solution is to deploy a new contract.
Also, if a new feature is added to the contract, the only way to do it is to deploy a new contract.
This is a problem for users, as they have to migrate to the new contract, and for developers, as they have to maintain multiple contracts.
To address this issue, several patterns have emerged, which are discussed later in the paper~\cite{chen2020maintaining,meisami2023comprehensive}.

\subsection{Gas and Storage in the Ethereum Virtual Machine (EVM)}\label{subsec:evm}
The EVM is a Turing-complete virtual machine that runs on every node of the Ethereum network and other EVM-based blockchains.
It provides a secure and isolated environment for smart contract execution~\cite{evm}.
Gas is the unit of computation in the EVM, and every operation performed by the EVM has a gas cost associated with it~\cite{gas-fees}.
The user pays this cost in the form of gas fees, to the miner/validator who executes the smart contract.
It is the miner's incentive to execute the smart contract and record the results on the blockchain, but also a way to prevent spam and denial of service attacks. Storage in the EVM is linked to the concept of gas, as every operation involving storage – whether it is writing new data or modifying existing data – incurs a gas cost. This cost is proportional to the storage resources consumed, reflecting the principle that the more network resources (like memory and storage space) a transaction uses, the more it needs to compensate the network.
Every smart contract has its own storage space, which is isolated and theoretically unlimited.

Smart contracts' storage utilizes a key-value store, with these key-value pairs referred to as storage slots.
A key for a storage slot is determined by the index of the slot, which is numbered contiguously from 0 to $2^{256}-1$.
A value is a 32-byte (or 256-bit) word, also called an item.
Data smaller than 32 bytes can be packed into a single slot, but if it is larger, the transaction is reverted~\cite{storage-layout}.
Some types of data are stored in multiple slots, such as arrays and mappings which are stored in multiple contiguous slots~\cite{mappings-arrays-storage}.
Strings are also stored in multiple slots, where one slot stores the length of the string, and the other slots store each 32-byte chunk of the string~\cite{bytes-string-storage}.
A \verb|struct|, which is a collection of variables, is stored in a single slot if it fits, otherwise, it is stored in multiple contiguous slots.
The gas cost of writing to storage increases with the number of slots written to.
This means that writing multiple variables, if they fit within a single slot, incurs the same gas cost as writing a single variable.

\section{Related Work}\label{sec:related-work}

\begin{figure}
   \centering
    \begin{subfigure}{\textwidth}
        \centering
    \includegraphics[width=.8\textwidth]{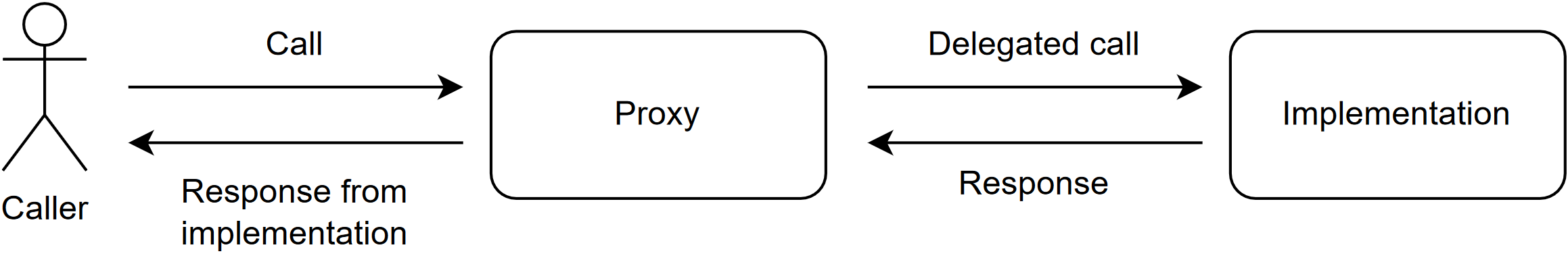}
    \caption{Proxy pattern diagram.}
    \label{fig:proxy-pattern-diagram}
    \end{subfigure}
    \hfill
    \begin{subfigure}{\textwidth}
        \centering
    \includegraphics[width=.8\textwidth]{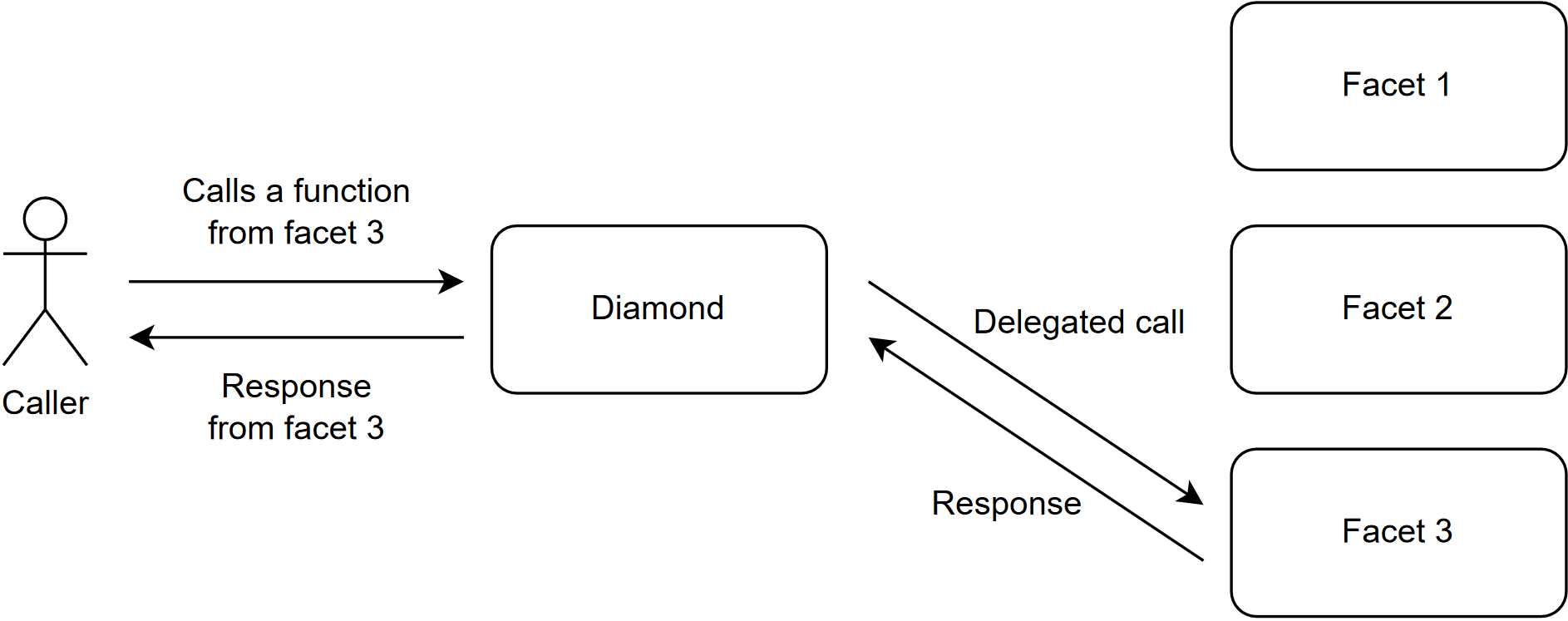}
    \caption{Diamond pattern diagram.}
    \label{fig:diamond-pattern-diagram}
    \end{subfigure}
    \caption{Illustration of the Proxy and Diamond patterns}
    \label{fig:illustration_patterns}
\end{figure}

Software engineering patterns provide well-tested solutions for frequently encountered application development use cases~\cite{alexander1979timeless}.
Standardized formatting such as the Alexandrian Form Format systematically include for each pattern the description, the forces or tradeoffs at stake, the benefits and drawbacks, and their main applications. A family of patterns focusing on blockchain patterns has recently emerged in the literature~\cite{xu2018pattern,six2022blockchain}.
The later include smart-contract related best practices such as upgradability patterns which refer to decentralized application maintenance strategies that can be applied for adding or updating features~\cite{chen2020maintaining}. Two recurring upgradability patterns are identified in the literature, namely the proxy and diamond patterns~\cite{marchesi2020design,ebrahimilarge,meisami2023comprehensive,qasse2023smart}.

The proxy pattern, pictured in Figure~\ref{fig:proxy-pattern-diagram} enables smart contract updates without changing the contract's address or requiring data migration\cite{proxy-pattern}.
It consists of two contracts, the user-interacted proxy and the logic-holding implementation. The proxy forwards calls to the implementation via delegated calls.
Shared storage ensures that if the implementation is updated, the storage persists.
To update the implementation, a new contract is deployed, and the proxy is directed to it, eliminating the need for user migration and preventing data loss. Compound, a key DeFi protocol, exemplifies this pattern through multiple upgrades, including Compound III~\footnote{\url{https://compound.finance/}}.
%Similarly, USDC, a widely used stable coin, leverages the proxy pattern for an upgrade to USDC 2.0.%, showcasing its adaptability in the evolving digital currency landscape.

The diamond pattern, depicted in Figure~\ref{fig:diamond-pattern-diagram} is a more upgradeable version of the proxy pattern.
This pattern solves the maximum contract size problem, which is a limitation of the proxy pattern. Indeed, logic can be separated into small contracts referred to as facets. 
A main contract, referred to as the implementation or diamond contract, points at the different facets to retrieve the applicable logic.
The diamond contract can be upgraded by adding, replacing or removing facets.
The diamond pattern is used in a diverse array of projects, as documented in Nick Mudge’s Awesome Diamonds repository\cite{awesome-diamonds}.
For example, Aavegotchi, a Non-Fungible Token (NFT) based gaming protocol, employs a single diamond pattern with eight distinct facets~\footnote{\url{https://www.aavegotchi.com/}}. %Each diamond and facet serves a specific purpose, contributing to the protocol's overall functionality.

A set of studies provide insight on upgradeable smart contract patterns.
Kannengiesser et al.\ conduct a study on key smart contract development challenges across various distributed ledger technology (DLT) protocols, including Ethereum, Hyperledger, and EOSIO~\cite{kannengiesser2021challenges}.
They highlight upgradability as a significant challenge and reference two upgradeable smart contract patterns, namely the diamond (referred to as the façade pattern) pattern and the proxy pattern.
However, the paper lacks an in-depth analysis of these patterns.
Two papers identify the proxy pattern but do not detail the forces, advantages, drawbacks, or gas costs considerations of this pattern~\cite{marchesi2020design,ebrahimilarge}.
There is no mention of the diamond pattern.
Two other works present a set of upgradeable smart contract patterns, including the proxy and diamond patterns~\cite{meisami2023comprehensive,qasse2023smart}.
However, the papers do not provide an in-depth analysis of the functional and non-functional properties of these patterns, nor gas costs behaviors.
Additionally, it is to note that two studies focus on gas cost efficiency strategies in smart contracts development.
Zarir et al.'s work focuses on transaction parameters rather than architectural design~\cite{zarir2021developing}.
The study by Di et al. identifies smart contract coding metrics impacting gas costs~\cite{di2022profiling}.
However, it does not specifically focus on upgradeable smart contracts.
These contracts possess unique functionalities like proxy pointers and proxy contract management.
In summary, a research gap is identified regarding a thorough study of the proxy and diamond patterns.
There is a lack of studies about gas costs analysis and formalization of the proxy and diamond patterns, especially using the Alexandrian form format.

\section{Methodology}\label{sec:methodology}

The methodology section of this paper details the approach employed for comparing the proxy and diamond patterns.
This involves the definition of a baseline scenario and the development of a gas consumption evaluation test bench.

\subsection{Protocol}\label{subsec:protocol}
\begin{figure}
    \centering
    \includegraphics[width=1\textwidth]{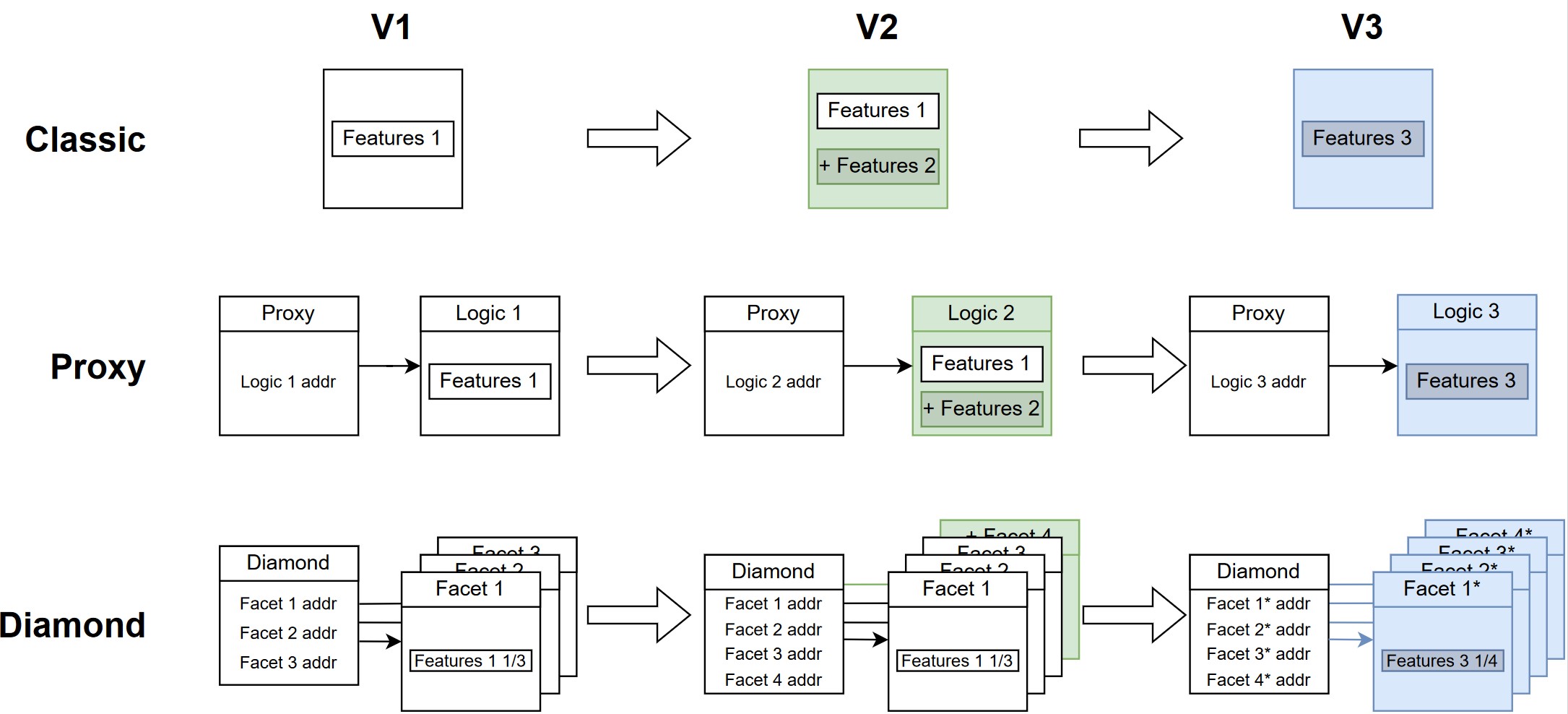}
    \caption{Evolution of the notarization application across versions by patterns.}
    \label{fig:upgrades-by-patterns-diagram}
\end{figure}

Smart contract deployments are one of the most expensive operations in terms of gas costs.
Therefore, assessing the deployment costs associated with various patterns is essential.
Additionally, upgrades often necessitate deploying additional smart contracts, making it crucial to compare the deployment costs of each subsequent upgrade.
The protocol leverages a file notarization scenario, a standard use of smart contracts to ensure the integrity of critical data such as diplomas or scientific workflows~\cite{badr2019permissioned,coelho2022blockchain,leible2019review}.
The scenario is implemented using the proxy and diamond patterns, as well as a reference monolithic non-upgradeable smart contract referred to as the classic pattern.
For each pattern, the initial deployment of the notarization smart contract is referred to as version 1.
The study then proceeds to two sequential upgrades of the application, a minor one (version 2) and a major one (version 3), to evaluate the gas costs.
Features are added as updates are made.
Figure~\ref{fig:upgrades-by-patterns-diagram} demonstrates the upgrades for each pattern.
Then the analysis turns to comparing the average gas cost of each function, for seeing which pattern is the less gas intensive when it comes to execution.
Each function is accompanied by a test case ran hundreds of times, to simulate real-world usage of this file notarization application over an extended period and to check the robustness of the code.

\subsection{Implementation}\label{subsec:implementation}

The code is written in Solidity, the main language used to develop decentralized application on EVM blockchains.
For the experiment, the Universal Upgradeable Proxy Standard (UUPS) proxy pattern is used because it is the recommended standard at the time of writing by OpenZeppelin~\cite{uups-proxy}.
For this study, the diamond pattern implemented by SolidState is used.
It utilizes Nick Mudge's gas-efficient Diamond 2 model~\cite{solidstate-diamond,nick-mudge-diamond-2} in a plug-and-play fashion: the developer only needs to import the diamond made by SolidState without any configuration needed.

\textbf{Version 1: Basic Notarization Application.}
In the first version of the notarization use case, files are represented by a mapping between their name and their hash.
The logic encompasses several functions.
The function \verb|addFile| is designed to notarize a file on the blockchain and modifies the state of the contract.
The \verb|getFileName| function is a view function that returns the name of the file.
Similarly, \verb|getFileHash| is a view function that provides the hash of the file.
Lastly, \verb|compareHash| is a pure function for comparing two hashes passed as parameters; while this operation is ideally performed off-chain, it is included here to demonstrate the gas cost of a pure function.

\textbf{Version 2: Updatable File.}
In case of file modifications, a function for updating it on the blockchain is needed.
The second version of this notarization application contains the \verb|updateFile| function, added on top of the previous version.
This function updates the file hash of a notarized file and so modifies the state of the contract.
This version can be seen as a minor update.

\textbf{Version 3: Access Control.}
For security reasons, access control is mandatory, so that only the owner of a file can modify or delete it.
This involves creating a \verb|File| structure containing the owner's address, the hash of the file's contents, the creation timestamp and the last modification timestamp.
The previous mapping is replaced by one between the file's name and its \verb|File| structure instance.
Then, the \verb|addFile| and \verb|updateFile| functions are modified to work with this new \verb|File| structure, and access control is added, where the new logic smart contract ensures that the caller interacts only with his own files.
This version can be seen as a major update.

\subsection{Unit Tests}\label{subsec:unit-tests}

As mentioned in section~\ref{subsec:protocol}, each function has a unit test that is run hundreds of times (704 iterations for the addFile function across all unit tests of version 3, for example).
It calls the function with random parameters (file name, hash, etc.) and checks that the result is correct.
This is done to simulate real-world usage of this file notarization application.
An expected result is computed before the function call, and the actual result is compared to it.
In the \verb|addFile| unit test, we expect the smart contract to add the file name and hash to its storage, using either the proxy or diamond pattern.
The \verb|updateFile| unit test should update the file hash in the contract's storage, while the \verb|deleteFile| test ensures the file's removal.
In the \verb|compareHash| test, the outcome should be true for matching hashes and false otherwise.
The \verb|getFileName|, \verb|getFileHash|, \verb|getFileOwner|, \verb|getFileCreatedAt|, and \verb|getFileLastModifiedAt| unit tests respectively verify the return of the file's name, hash, owner address, creation timestamp, and last modification timestamp.
Finally, the \verb|getFileDetails| test checks for the return of all the previously mentioned file properties.

Testing each single function is important to ensure that the code is robust and that it does not break when upgrading the smart contract.
These tests are also conducted to compare the gas costs across various patterns, specifically examining each type of function, including pure, view, and state-modifying functions.
By testing pure functions, the gas costs of computations without storage access can be assessed and compared across patterns.
View functions are tested to evaluate the gas costs of computations with storage access.
Finally, state-modifying functions are tested to assess the gas costs of storage writes.

Smart contracts used for this study are developed using Foundry~\cite{foundry}.
Foundry plays a critical role in this testing process by autonomously simulating an Ethereum Virtual Machine (EVM) blockchain environment. During test execution, it deploys the contracts and carries out the testing scenarios within this emulated setting, utilizing the default configuration of this local blockchain.
The entire source code, tests and results here are available in the accompanying source code repository~\footnote{\url{https://anonymous.4open.science/r/proxy-diamond-patterns-gas-analysis}}.

%This toolkit is chosen because it compiles and executes faster than other well-known frameworks such as Truffle and Hardhat. Moreover, It eliminates the need for an additional programming language for writing tests and scripts, unlike Truffle and Hardhat with JavaScript, and generates easily gas reports per smart contract function. 

\subsection{Results Retrieval}\label{subsec:results-retrieval}

Through these unit tests, Foundry produces gas reports that provide insights into the gas consumption for each function, alongside the gas costs associated with deployment and the size of the contracts.
The cost of function calls is quantified in gas units, and the contract size is measured in bytes.

These gas cost results for each function are derived by summing the gas costs associated with every operation performed within the function.
Each operation, also known as opcode, has a fixed gas cost, which is the same for all patterns.
Those instructions are described in the Ethereum Yellow Paper~\cite{wood2014ethereum}.
For example, the \verb|SLOAD| opcode, which reads a value from storage, has a fixed cost of 100 for warm access, and 2100 gas for cold access.
Then, the deployment cost is the sum of several components.
First, the \verb|TRANSACTION| opcode incurs 21,000 gas units, representing the base cost of every transaction on the EVM\@.
Additionally, there is the \verb|CREATE| opcode, costing 32,000 gas units, which is used for creating a new contract.
Next, the cost related to the bytecode includes 4 times the number of 0 bytes and 16 times the number of non-zero bytes.
Furthermore, 200 gas units are added for every byte of the contract's size.
Finally, if a constructor function is present, its cost is also included in the deployment cost.

In addition, the framework furnishes complete stack traces of the calls, detailing the gas consumption for each operation.
These traces are utilized to construct charts that capture the gas cost of each function call.
This level of detail supplements gas reports, which typically provide only a summary of costs, including the minimum, average, median, and maximum values.

\section{Evaluation}\label{sec:evaluation}

This section presents the gas costs evaluation of the proxy and diamond smart contract patterns, against a baseline built using the classic pattern in the context of an app deployment and upgrade~\footnote{All results can be found here: \url{https://anonymous.4open.science/r/proxy-diamond-patterns-gas-analysis/data}}.

\begin{figure}
   \centering
    \begin{subfigure}{.48\textwidth}
        \centering
        \includegraphics[width=\textwidth]{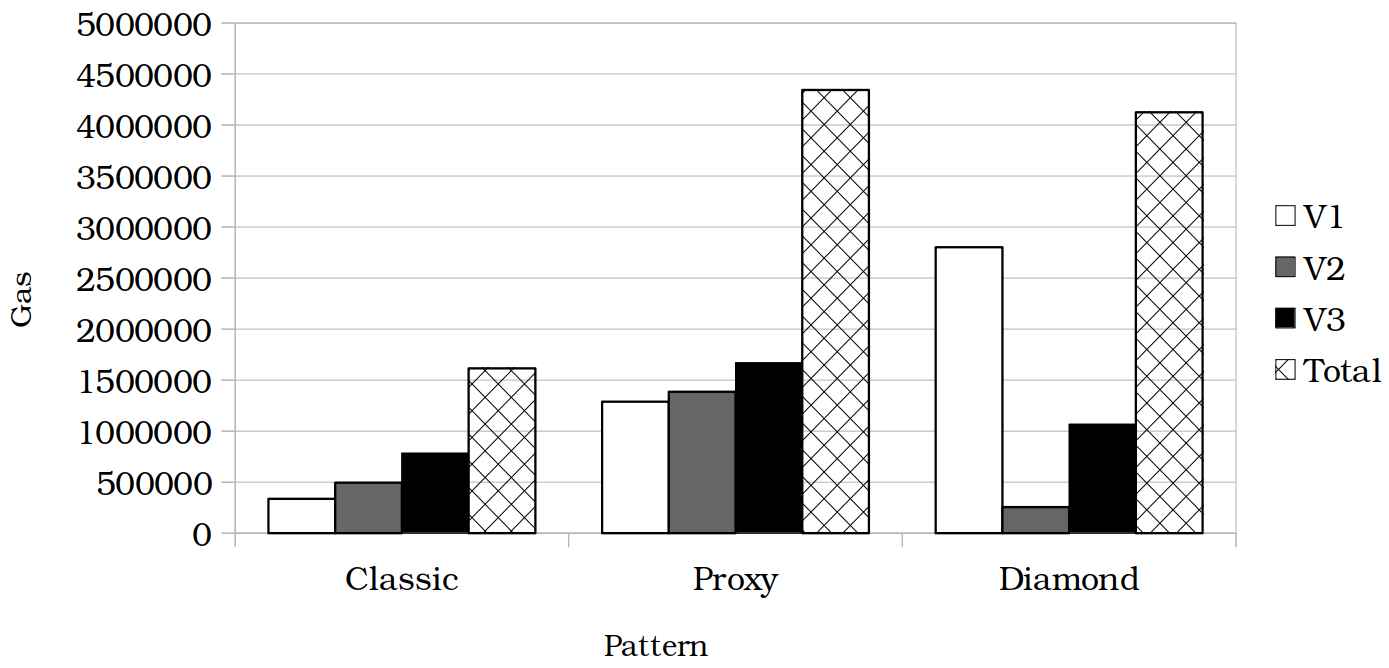}
        \caption{Deployment cost by pattern and version.}
        \label{fig:deployment-cost-pattern-version}
    \end{subfigure}
    \hfill
    \begin{subfigure}{.48\textwidth}
        \centering
        \includegraphics[width=\textwidth]{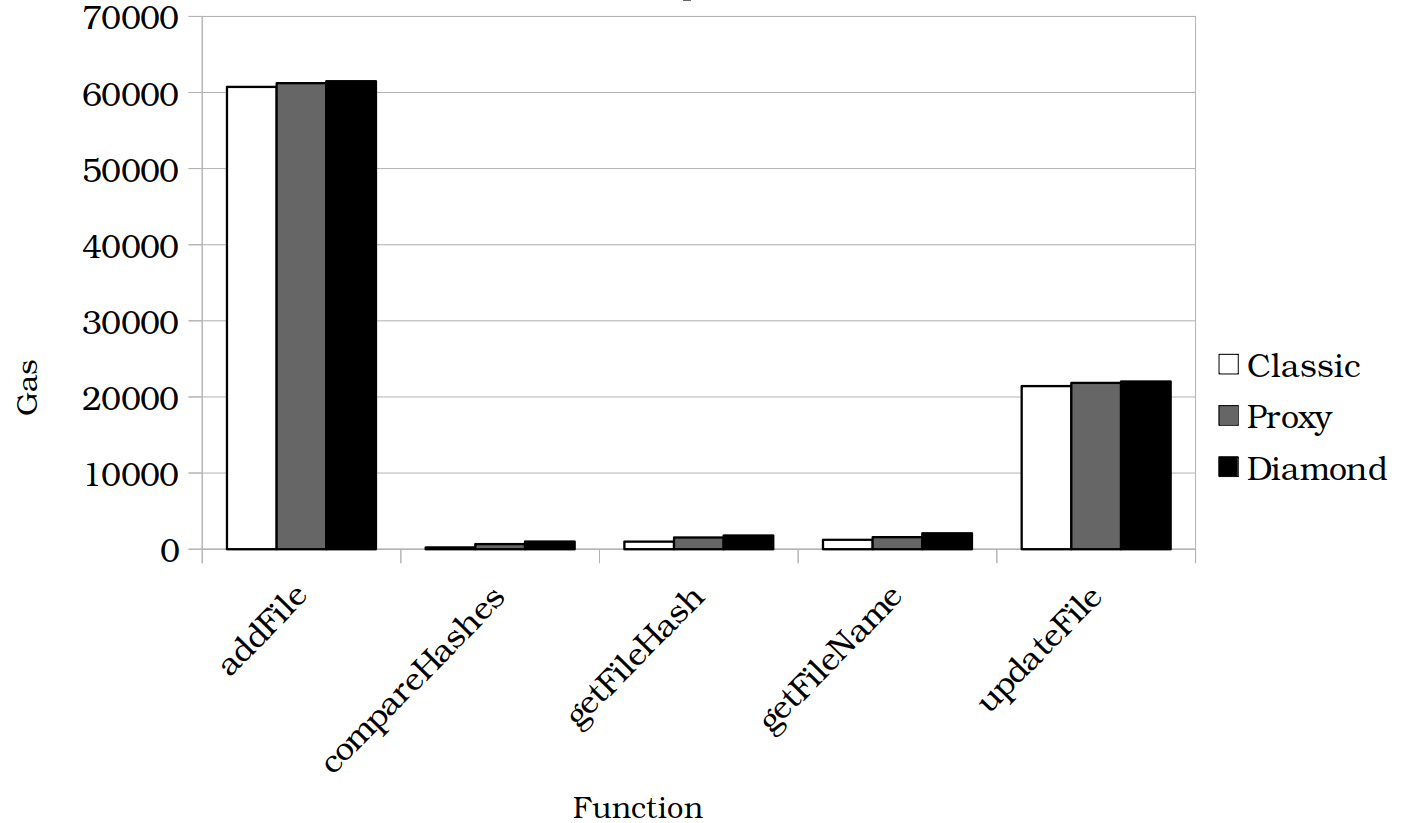}
        \caption{Average function cost by pattern for version 2.}
        \label{fig:average-v2-function-cost-pattern-chart}
    \end{subfigure}
    \caption{Evaluation results}
    \label{fig:evals}
\end{figure}

\subsection{Gas Cost During Smart Contract Deployment and Upgrades}\label{subsec:gas-cost-during-upgradeable-smart-contract-deployment-and-upgrades}

The primary aim is to compare the gas costs during the deployment (version 1) and upgrades (version 2 and 3) of the file notarization application implemented with the three different patterns.
For the classic and proxy pattern, a new contract is deployed for each version.
For the proxy, the proxy pointer is updated in the smart contract implementation.
For the diamond pattern, in version 2, only facet is changed, and in version 3, all facets are changed.

Figure~\ref{fig:deployment-cost-pattern-version} presents the deployment costs, in gas units, of each version by pattern.
The cost of deployment increases with each version, except for the diamond pattern, where the initial cost of deployment is significant.
In all, the classic pattern requires just 1,614,545 units of gas, while the proxy and diamond patterns consume around 2.6 times as much, at 4,343,104 and 4,123,977 units of gas respectively.
The classic pattern appears the most gas efficient.
The increased consumption at deployment of the proxy and diamond patterns relates to the need to deploy more contracts compared to the classic pattern: two for the proxy pattern, and four for the diamond one.

\subsection{Gas Cost During Smart Contract Execution}\label{subsec:gas-cost-during-upgradeable-smart-contract-execution:-study-of-all-functions}

The analysis now shifts to comparing the average gas cost of functions to determine the most gas efficient execution pattern. Gas cost behaviors in versions 1 and 3 align with online gas reports. Notably, \verb|addFile| and \verb|updateFile| functions are the costliest, while remaining functions incur comparatively lower gas costs. Across the three patterns, the proxy and diamond patterns exhibit slightly higher expenses than the classic pattern. Notable cost discrepancies include \verb|compareHashes| being significantly cheaper than \verb|addFile| in the classic pattern. This difference is attributed to \verb|addFile| requiring two storage writes while other functions mainly perform storage reads. The classic pattern avoids delegation to other contracts, unlike the proxy and diamond patterns, resulting in additional costs. Furthermore, incurring more operations when invoking a diamond adds to its expense compared to a proxy. Getters in the classic pattern involve state lookups, costing 2100 gas for cold storage loads, while \verb|compareHashes| is a \verb|pure| function, devoid of state lookups or modifications, resulting in lower gas costs.

In reviewing the data obtained from the comparative analyses, the deployment phase exhibits the most significant variation among the different patterns, primarily due to the fluctuating number of smart contracts deployed.
For this metric, the diamond pattern is the cheapest, especially for minor upgrades.
During execution, the gas costs are similar across patterns, though there is a marginal escalation in costs when transitioning from the classic architecture to the proxy pattern, and subsequently from the proxy to the diamond pattern.
This increase in gas costs is due to the additional operations required to delegate calls to the logic contract(s) in the proxy and diamond patterns.
\section{Discussion}\label{sec:discussion}

This section summarizes and consolidates the proxy and diamond patterns following the Alexandrian form format proposed by C. Alexander~\cite{alexander1979timeless}. A decision model is proposed to help developers choose between both patterns.

\subsection{Proxy Pattern Outline}\label{subsec:proxy-pattern-outline}
\begin{itemize}
    \item \textbf{Summary:} The proxy pattern facilitates upgradability in smart contracts through a proxy contract pointing to the latest version of a logic contract.

    \item \textbf{Context:} A smart contract must be upgraded due to evolving requirements and potential improvements~\cite{buterin2017ethereum}.

    \item \textbf{Problem:} Traditional smart contracts lack the ability to be updated without manual storage migration, posing challenges in addressing vulnerabilities, enhancing functionality, and adapting to changing circumstances.

    \item \textbf{Forces (tradeoffs):} The problem requires balancing the following forces: (i) \textit{Immutability vs. Upgradability.} Smart contracts on blockchain platforms are traditionally immutable once deployed; (ii) \textit{Gas Costs vs. Flexibility.} Minimizing gas costs while providing flexibility for contract upgrades; (iii) \textit{Trust vs. Transparency.} Establishing trust in the upgrade process while maintaining transparency.
    \item \textbf{Solution:} The smart contract proxy pattern introduces a proxy contract as an intermediary layer.
    This proxy delegates calls to a logic contract, allowing seamless upgrades by deploying new logic contracts and updating the proxy to point to the latest version, hence allowing for dynamic updates.

    \item \textbf{Consequences:}
    \begin{itemize}
        \item Benefits:
        \begin{itemize}
            \item \textit{Upgradability.} Allows upgrades without the need to change the contract address nor requiring data migration;
            \item \textit{Simplest Upgradeable Pattern.} The upgrade process consists in deploying a new logic contract and a proxy update pointing to it with a proxy administration function.
        \end{itemize}
        \item Drawbacks:
        \begin{itemize}
            \item \textit{Compatibility Maintenance.} Requires  consistent function selectors and storage layouts;
            \item \textit{Limited Direct Function Visibility.} Logic functions visibility only accessible in the documentation;
            \item \textit{Storage Collision.}  It requires careful consideration of storage layout to avoid storage overlap between the proxy and the logic contract as if both contracts use the same storage slot, it can lead to data loss. A convention is to use namespaced storage layouts for naming struct holding storage variables~\cite{eip-7201}.
            \item \textit{Function Selector Clash.} Different functions having the same selector can override each other~\cite{function-selector-clashing}. This requires careful naming of function selectors as the Solidity compiler cannot detect function selector clashes between the proxy and logic contracts due to cross-contract interactions.%, so that no function is overwritten.
            %Since proxy patterns require cross-contract interaction, the Solidity compiler cannot detect these clashes: it is the developer's responsibility to avoid them.
            
        \end{itemize}
    \end{itemize}

    \item \textbf{Related patterns:} Diamond pattern.

    \item \textbf{Known uses:}
    Two standardized implementations are proposed in the Ethereum Improvement Proposal \textit{EIP-897} and in OpenZeppelin's  smart contract development framework~\cite{meisami2023comprehensive,amri2023review}. 
   \textit{Compound} and \textit{USDC}, respectively a DeFi protocol and a stable coin, both implement the proxy pattern for upgrades. 
\end{itemize}

\subsection{Diamond Pattern Outline}\label{subsec:diamond-pattern-outline}

\begin{itemize}
    \item \textbf{Summary:} employing multiple implementation contracts to balance contract size, maintainability and versioning;
    \item \textbf{Context:} Need for a solution to improve maintainability in large contracts.
    \item \textbf{Problem:} Traditional approaches face challenges in managing contract size and versioning effectively.
    \item \textbf{Forces (tradeoffs):} The problem requires balancing the following forces: (i) \textit{Contract Size vs. Modularity} Balancing the need for compact contract sizes with the demand for modular, well-organized code structures; (ii) \textit{Maintainability vs. Simplicity} Achieving improved maintainability without
    introducing unnecessary complexity; (iii) \textit{Scalability vs. Consistency} Scaling smart contract applications requires accommodating versioning and updates.
    \item \textbf{Solution:} The diamond pattern introduces a structure where a proxy can point to multiple logic contracts.
    It involves the deployment of all contracts (diamond and facets), retrieving facets function selectors, and leveraging both to implement a diamond cut.
    \item \textbf{Consequences:}
    \begin{itemize}
        \item Benefits:
        \begin{itemize}
            \item \textit{Better upgradability.} Possibility to deploy smaller contracts during upgrades or updates to already deployed facets without requiring address change or data migration; 
            \item \textit{Modularity.} Code reusable across multiple contracts. A facet can be used in multiple diamonds;
            \item \textit{Contract size.} Thanks to a modular structure, it can theoretically support an infinite number of facets. Therefore, the whole smart contract system has no size limit;
            \item \textit{Cheaper minor upgrades.} Most of the time, only one facet is updated, so only small contracts are deployed for low gas costs;
            \item \textit{Shorter compilation time.} Only modified facets need to be compiled, so for the same logic code, the compilation time is shorter than for the classic pattern and proxy pattern.
        \end{itemize}
        \item Drawbacks:
        \begin{itemize}
            \item \textit{Implementation Complexity.} A more complex structure compared to the classic and proxy patterns, and a lack of supporting libraries.
            \item \textit{Complexity in managing multiple logic contracts.} Managing multiple logic contracts require careful consideration during upgrades. It requires developers to manage the diamond storage manually because of the multiple implementation contracts.
            \item \textit{Limited Direct Function Visibility.} Like the proxy pattern, users depend on documentation to identify callable functions.
            \item \textit{Storage collision risks} similarly to Proxy pattern
            \item \textit{Function selector clash} similarly to Proxy pattern
        \end{itemize}
    \end{itemize}
    \item \textit{Related patterns:} Proxy pattern.
    \item \textit{Known uses:} \textit{Aavegotchi,} a NFT-based gaming protocol; \textit{GeoWeb,} a dApp managing digital land property rights using NFTs.
\end{itemize}

\subsection{Choosing between the Proxy and Diamond Patterns}\label{subsec:choosing-between-proxy-and-diamond-patterns}

\begin{figure}
    \centering
    \includegraphics[width=\textwidth]{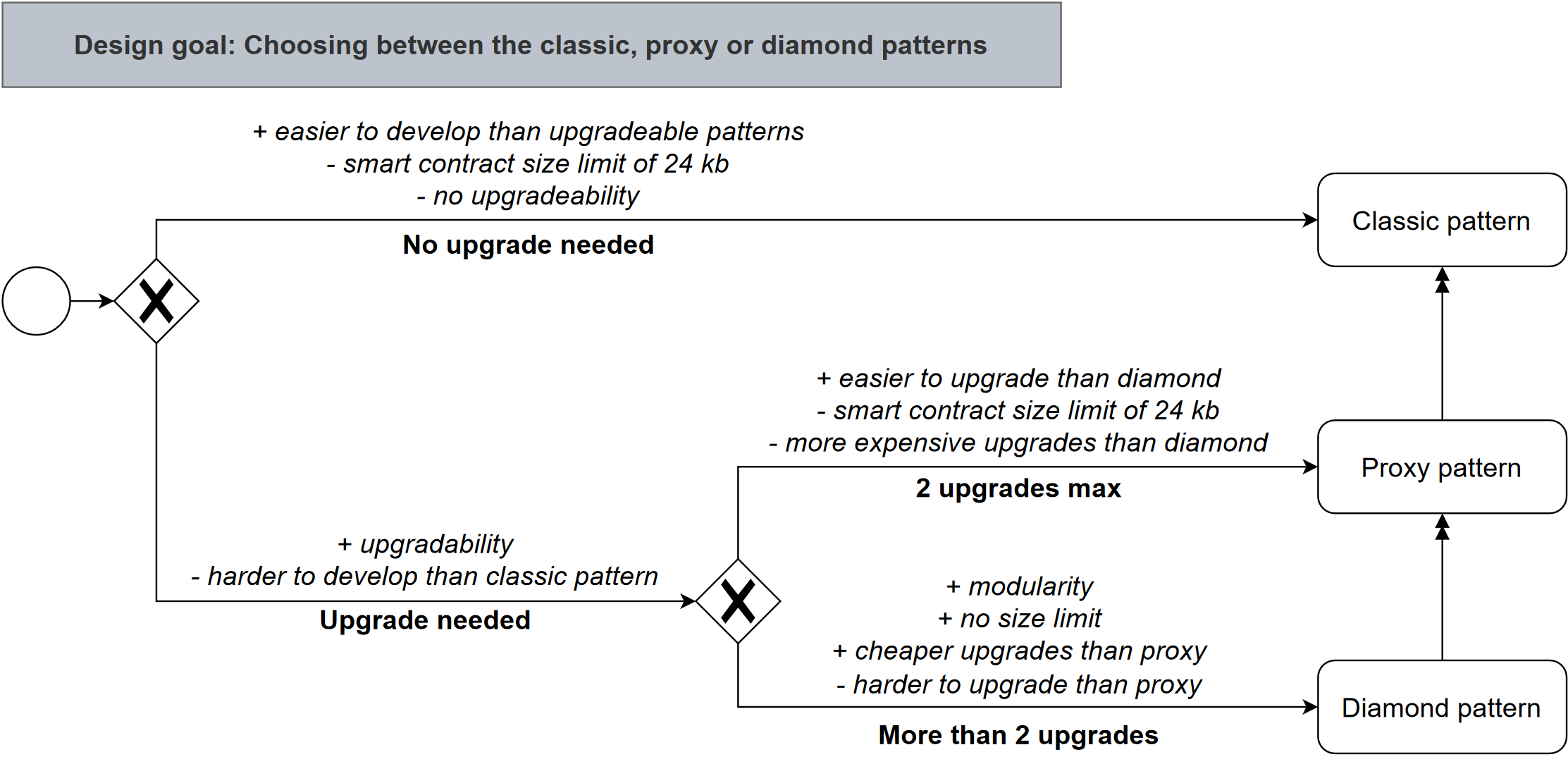}
    \caption{Decision model for upgradeable smart contract pattern usage.}
    \label{fig:decision-model-patterns}
\end{figure}

Figure~\ref{fig:decision-model-patterns} proposes a decision model to orient between a proxy or a diamond pattern in the design stage of an upgradeable smart contract.
It follows the design goal decision model introduced by Xu where decisions are modeled using a logical BPMN flow~\cite{xu2021decision}.
More precisely, arrows, logical gateways, and functional and non-functional properties orient the decision path.

The choice of smart contract pattern largely depends on the need for upgradability.
For scenarios without upgradability needs, the classic pattern is preferable due to its straightforward development process, relying on contract inheritance and library imports.
However, upgrades in the classic pattern often involve complex and resource-intensive data migration.
This pattern also poses challenges in communicating new contract addresses to users, potentially affecting the user experience.
For extensive upgradeable features, the diamond pattern is recommended.
Its modular nature allows for easy addition or removal of facets, reducing compilation time.
However, it is less cost-effective initially compared to the proxy pattern and requires in-depth knowledge of smart contract storage and facet-library management.
The proxy pattern is advised for limited code sizes or infrequent upgrades.
It simplifies development and integrates easily with libraries like OpenZeppelin’s.
This pattern enhances upgradability by separating logic and state, reducing the need for data migration.
But it offers less flexibility and modularity compared to the diamond pattern and demands careful consideration to maintain compatibility across versions.

In the end, while the classic pattern excels in execution ease, the proxy and diamond patterns provide a more manageable framework for upgrades, simplifying contract interactions for users during updates.
The diamond pattern is more suitable for extensive upgrades, while the proxy pattern is recommended for limited upgrades and simpler development.

\section{Conclusion}\label{sec:conclusion}

In conclusion, this comprehensive study delves into the intricacies of upgradeable smart contract design patterns—a critical facet of contemporary blockchain applications.
The comparative analysis specifically focuses on the gas costs associated with deploying, using, and upgrading decentralized applications using two prominent upgradeable patterns: the proxy and diamond patterns.

Each pattern unfolds with distinct strengths and weaknesses, delineating its applicability across diverse scenarios.
The classic pattern implies an unpractical and costly approach to smart contract upgrades because of the need of manual data migration.
The proxy pattern offers the simplest solution for upgradability, but security concerns such as storage collisions and function selector clashes remain.
This demands a developer's careful attention and thus results in a more challenging pattern to utilize.
The diamond pattern is the most complex of the three, but it offers the most flexibility and maintainability thanks to its modularity.
Moreover, the diamond pattern is the most gas efficient when it comes to doing more than two upgrades.
This is because the diamond pattern does not need to deploy long contracts, but only small facets.
Finally, this pattern is the most scalable because it does not have a contract size limit. Despite these considerations, real-world implementations in projects like Compound, USDC, GeoWeb, and Aavegotchi underline the use of the proxy and diamond patterns as facilitators for flexible, upgradeable, and scalable smart contracts.

To generalize these initial findings, the study advocates for extending experiments beyond the notarization scenario used as a comparison baseline for this paper.
Essential to this endeavor is the necessity for additional experiments encompassing diverse blockchain networks and pattern libraries to extrapolate and validate the findings.
For future work, a replication of the study on alternative upgradeable patterns would also provide a more comprehensive understanding of upgradeable smart contract patterns.

    \bibliographystyle{splncs04}
    \bibliography{references}

\end{document}